\documentclass[letterpaper,10pt,conference]{cssconf}

\IEEEoverridecommandlockouts                              
\usepackage{amsmath}
\usepackage{theorem}
\usepackage{amsfonts}
\usepackage{graphicx}
\usepackage[colorlinks=false, hidelinks]{hyperref}
\usepackage{xcolor}
\usepackage{cite}
\usepackage{footmisc}

\allowdisplaybreaks

\newcommand{\cN}{{\mathcal N}}

\newcommand{\oo}{{\mathsf o}}
\newcommand{\s}{{\mathsf s}}

\newcommand{\R}{{\mathbb R}}

\newtheorem{Theorem}{Theorem}
\newtheorem{Lemma}[Theorem]{Lemma}

\newtheorem{Assumption}{Assumption}
{\theorembodyfont{\rmfamily} }
{\theorembodyfont{\rmfamily} }

\colorlet{Darkred}{blue!90!white} 
\colorlet{Darkgreen}{orange!85!black} 
\newcommand{\cdr}[1]{\textcolor{Darkred}{#1}}
\newcommand{\cdg}[1]{\textcolor{Darkgreen}{#1}}

\newcommand{\mathbookmark}[2]{\texorpdfstring{#1}{#2}}

\title{\bf Adaptation of Parameters in Heterogeneous Multi-agent Systems}

\author{Hyungbo Shim, Jin Gyu Lee, and Brian D.~O.~Anderson
	\thanks{H.~Shim and J.G.~Lee are with ASRI and the Department of Electrical and Computer Engineering, Seoul National University, Korea (email: \url{hshim@snu.ac.kr} and \url{jingyu.lee@snu.ac.kr}).}
	\thanks{B.D.O.~Anderson is with the School of Engineering, Australian National University, Acton, ACT 2601, Australia (email: \url{brian.anderson@anu.edu.au}).}
	\thanks{This work was supported by the National Research Foundation of Korea (NRF) grant funded by the Korea government (MSIT) (No.~RS-2022-00165417).}
}

\begin{document}

\maketitle

\begin{abstract}
This paper proposes an adaptation mechanism for heterogeneous multi-agent systems to align the agents' internal parameters, based on enforced consensus through strong couplings. Unlike homogeneous systems, where exact consensus is attainable, the heterogeneity in node dynamics precludes perfect synchronization. Nonetheless, previous work has demonstrated that strong coupling can induce approximate consensus, whereby the agents exhibit emergent collective behavior governed by the so-called blended dynamics. Building on this observation, we introduce an adaptation law that gradually aligns the internal parameters of agents without requiring direct parameter communication. The proposed method reuses the same coupling signal employed for state synchronization, which may result in a biologically or sociologically plausible adaptation process. Under a persistent excitation condition, we prove that the linearly parametrized vector fields of the agents converge to each other, thereby making the dynamics asymptotically homogeneous, and leading to exact consensus of the state variables.
\end{abstract}


\section{Introduction}

Examples of collective phenomena include bird flocking, fish schooling, and the synchronization of beating cells. Such phenomena can be explained by the concept of state consensus in groups of dynamical systems. Early studies on multi-agent systems primarily focused on the case where all agents have identical dynamics. However, attention soon shifted toward heterogeneous multi-agent systems, for which exact consensus is fundamentally unattainable unless every agent shares the same internal model. In particular, differences in node vector fields prevent identical integration, even when all agents start from the same initial condition \cite{wieland2011,kim2011}.

Consider for instance a network of \(N\) agents described by
\begin{equation}\label{eq:hmas}
\dot x_i = f_i(x_i,t) + k \sum_{j \in \cN_i} (x_j - x_i), \qquad i = 1, \cdots, N,
\end{equation}
where $x_i \in \R^n$ denotes the state of agent \(i\), \(\cN_i\) represents the set of neighboring agents of agent \(i\), and \(k\) is a constant coupling gain. The function \(f_i\) is a locally Lipschitz (and possibly time-varying) vector field that may vary from one agent to another (hence the subscript \(i\)). Were consensus to be achieved,  the coupling term \(k \sum_{j \in \cN_i} (x_j - x_i)\) would vanish, but any discrepancy of the \(f_i\) would remain, thereby preventing exact consensus. 

Despite this inherent limitation implying that exact consensus is unachievable, it was shown in \cite{kim2015a,panteley2017} that strong coupling---often achieved by choosing a large \(k\)---can enforce approximate consensus. More precisely, if the communication graph is undirected and connected, and if the so-called {\em blended dynamics}
\begin{equation}\label{eq:blendeddynamics}
\dot s = \frac{1}{N} \sum_{i=1}^N f_i(s,t)
\end{equation}
is contractive (in the sense similar to Assumption \ref{ass:bd} below), then for any \(\epsilon > 0\) there exists a threshold \(k^* > 0\) such that for every \(k \geq k^*\) the trajectories \(x_i(t)\) satisfy\footnote{This result is global if the \(f_i\)'s are globally Lipschitz; otherwise, it is semi-global.}
\begin{equation}\label{eq:error}
\limsup_{t \to \infty} \|x_i(t) - s(t)\| \;\le\; \epsilon, 
\quad i = 1, \cdots, N,
\end{equation}
where \(s(t)\) is the solution of \eqref{eq:blendeddynamics} with \(s(0) = \frac{1}{N}\sum_{i=1}^N x_i(0)\) \cite{kim2015a,lee2020a}.

\begin{figure}[t]
\centering
\includegraphics[width=0.48\textwidth]{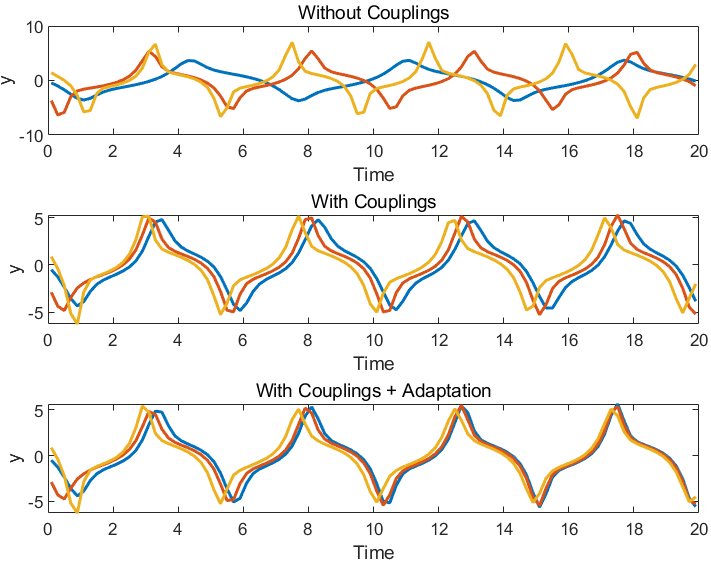}
\caption{Trajectories of the first state variable from three heterogeneous Van der Pol oscillators. (Top) Without coupling, they evolve differently. (Middle) With strong coupling, they nearly synchronize but not exactly. (Bottom) With strong coupling and the proposed adaptation algorithm, 
the parameters of the node dynamics achieve consensus, yielding a common vector field, and thus, exact synchronization over time. An animated simulation is available at \url{https://youtu.be/CVQxZgb5Pro} and the codes are at \url{https://github.com/hyungbo/vfconsensus}.}
\label{fig:1}
\end{figure}

Figure~\ref{fig:1} illustrates this behavior using three heterogeneous Van der Pol oscillators by showing one of their state variables. In the absence of coupling (\(k = 0\)), each oscillator exhibits distinct behavior (top subplot). With coupling (\(k > 0\)), approximate synchronization is observed (middle subplot). Notice that the synchronized behavior does not replicate any individual oscillator’s dynamics from the top subplot; rather, it represents an emergent behavior governed by the blended dynamics \eqref{eq:blendeddynamics} \cite{lee2018}.

This observation supports the idea that {\em seeking to enforce consensus in heterogeneous dynamical systems produces emergent behavior}, and indeed, even if exact consensus is \textit{not} reached but only approximate consensus is achieved, such emergent behavior arises.
Moreover, this general conclusion can be exploited in various engineering applications, such as distributed optimization \cite{lee2022} and distributed state estimation \cite{kim2020}. Interested readers may refer to \cite{lee2022b}. In these applications, heterogeneity is intentionally introduced as a design tool.

On the other hand, not all heterogeneity is intentional. Perturbations, manufacturing imperfections, and wear-and-tear during operation can also result in heterogeneous parameters that are typically undesirable and thus viewed as factors to overcome. Of course, strong coupling alone can in principle reduce the steady-state consensus error \eqref{eq:error} arbitrarily by increasing \(k\). Nonetheless, if additionally the individual internal parameters can be gradually adjusted so that they converge to identical values, then exact consensus can be achieved, without necessarily resorting to very large coupling gains. This approach is illustrated in the bottom subplot of Fig.~\ref{fig:1}.

Motivated by this idea, we propose a form of parameter update law that aligns the internal parameters of heterogeneous agents. More specifically, we consider the case where the vector field \(f_i\) is linearly parameterized as
\begin{equation}\label{eq:sys0}
\dot x_i = \psi(x_i,t) \,\theta_i + \psi_o(x_i,t) + k \sum_{j \in \cN_i} (x_j - x_i)
\end{equation}
where \(\theta_i \in \R^p\) is the parameter vector of agent \(i\), \(\psi: \R^n \times \R \to \R^{n \times p}\) is a given regressor, and \(\psi_o: \R^n \times \R \to \R^n\) represents the common part of \(f_i\). The proposed parameter update law is
\begin{equation}\label{eq:update}
\dot \theta_i = g \,\psi^T(x_i,t) \cdot k \sum_{j \in \cN_i} (x_j - x_i),
\end{equation}
where \(g\) is a small positive constant indicating slow evolution of $\theta_i$. 
If direct communication of the parameters \(\theta_i\) among agents were allowed, one might simply employ an average-consensus update:
\[
\dot \theta_i = \sum_{j \in \cN_i} (\theta_j - \theta_i).
\]
However, in order to maintain consistency with the biology/sociology motif, our approach crucially avoids \textit{explicit} parameter sharing. For example, if children in a kindergarten gradually develop similar behaviors over time, it is more plausible that they do so by spending time together (thereby synchronizing their external behavior) rather than by explicitly sharing their “internal parameters.” In \eqref{eq:update}, the same coupling signal as in \eqref{eq:sys0} is recycled for parameter adaptation, eliminating the need for an additional communication channel.
In this sense, the agent model \eqref{eq:sys0} and \eqref{eq:update} may explain the adaptation behavior of a complex swarm.

The main objective is to identify appropriate conditions for all \(\theta_i\) to tend to the same value as \(t \to \infty\) so that the agents’ vector fields effectively achieve consensus. 
Under such circumstances, the states \(x_i\) no longer just approximately synchronize---they converge to exact consensus.

To the best of the authors' knowledge, adaptation of vector fields in multi-agent systems has rarely been explored, and therefore, we set the goal of this conference paper to convey the core idea in a clear and accessible manner. 
For this, we adopt the following simplifying assumptions: (i) the function \(\psi_o\) is identically zero, (ii) the system state dimension is \(n = 1\), and (iii) the function \(\psi\) is continuously differentiable. 
These simplifications allow us to focus on the fundamental mechanisms underlying the proposed adaptation strategy, and their relaxation will be reported in a journal version.

In this paper, we use the usual Euclidean vector norm and its induced matrix norm, both of which are denoted by $\|\cdot\|$.
To aid readability, frequently used variables and constants are highlighted in color when they are introduced for the first time.

\section{Main Result}

Let us consider a group of agents described by the following dynamics for \( i = 1, \dots, N \):
\begin{align}
	\dot x_i &= \psi(x_i,t) \theta_i + k \sum_{j \in \cN_i} (x_j - x_i), & & x_i \in \mathbb{R}, \label{eq:mas} \\
	\dot \theta_i &= g \psi^T(x_i,t) \cdot k \sum_{j \in \cN_i} (x_j - x_i), & & \theta_i \in \mathbb{R}^p, \label{eq:adaptation_law}
\end{align}
where \cdg{\(k\)} and \cdg{\(g\)} are positive constants (typically large and small respectively, but whose values are to be discussed), and \(\cN_i\) denotes the set of neighbors of agent \(i\), as determined by the communication graph \(\mathcal{G}\). The graph \(\mathcal{G}\) and the function \(\psi\) are assumed to satisfy the following conditions.

\begin{Assumption}\label{ass:graph}
The communication graph \(\mathcal{G}\) is undirected and connected.
\end{Assumption}

Some form of connectedness is self-evidently a requirement if consensus, approximate or exact, is desired. The assumption of undirectedness is made for convenience because, with the directed graph having a rooted spanning tree, the form of the blended dynamics becomes more complicated, being a weighted sum of individual node dynamics.

\begin{Assumption}\label{ass:psi}
The function \(\psi: \R \times \R \to \R^{1 \times p}\) is continuously differentiable and uniformly bounded (together with its first partial derivatives) with respect to \(t\); that is, there exists a continuous function \(\rho\) such that, for all \(t\) and \(s\),
$$\max\left\{ \|\psi(s,t)\|, \left\| \frac{\partial \psi}{\partial s}(s,t) \right\|, \left\| \frac{\partial \psi}{\partial t}(s,t) \right\| \right\} \le \rho(\|s\|).$$
\end{Assumption}

Assumption \ref{ass:psi} prevents the case of unbounded vector fields over time.

The blended dynamics of \eqref{eq:mas} is defined as
\begin{equation}\label{eq:bd0}
\dot s = \frac1N \sum_{i=1}^N \psi(s,t) \theta_i(t), \qquad \cdg{s(0)} = \frac{1}{N} \sum_{i=1}^N x_i(0)
\end{equation}
and an associated condition typically involves some form of stability of the blended dynamics \eqref{eq:bd0}.
However, since the evolution of $\theta_i(t)$ along \eqref{eq:adaptation_law} is not known {\it a priori}, we impose more easily verifiable conditions on the following dynamics of \cdg{$\hat s$}:
\begin{equation}\label{eq:bd}
\dot {\hat s} = \psi(\hat s,t) \cdot \theta_0^*
\end{equation}
with a constant \cdg{$\theta_0^*$}.

\begin{Assumption}\label{ass:bd}
There is a compact set $\cdg{\Theta_0^*} \subset \R^p$ such that, for each $\theta_0^* \in \Theta_0^*$, the system \eqref{eq:bd} is contractive (with a state-independent metric)\footnote{\label{footnote1}A system $\dot x = f(x,t)$, $x \in \R^n$, is contractive with a state-independent metric if $\exists$ a positive definite matrix $M$ and a constant $c>0$ such that $M \frac{\partial f}{\partial x}(x,t) + \left(\frac{\partial f}{\partial x}(x,t)\right)^T M \le - 2c I_n$, $\forall x,t$. In general, contractivity is defined with a state-dependent metric $M(x)$ \cite{lohmiller1998}.}; that is, there exists $\cdg{c}>0$ such that
\begin{equation}\label{eq:c}
\frac{\partial \psi}{\partial s}(s,t) \cdot \theta_0^* \le -c, \qquad \forall s,t \in \R, \; \forall \theta_0^* \in \Theta_0^*.
\end{equation}
\end{Assumption}

The role of Assumption \ref{ass:bd} is to ensure with sufficiently large $k$ an approximate consensus, despite the heterogeneity \cite{lee2020a}.
Indeed, without some form of stability of the blended dynamics, (approximate) consensus cannot be sustained under strong coupling in general \cite{lee2020a}.

While exact consensus of state variables is an outcome of the adaptation of parameters for agreement, it will be seen that the proposed adaptation mechanism is induced by approximate synchronization of state variables.
Synchronization-induced adaptation may be considered consistent with the conclusion of \cite{Azzi2017}, which asserts that the parameters of the circadian clock are adjusted through synchronization with external light and darkness.
We will clarify how synchronization induces adaptation through the {\em persistent excitation} condition on \(\psi\) stated below.

\begin{Assumption}\label{ass:pe}
There are positive \(c_1\), \(c_2\), \(\tau\), and $M_x$ such that, for each $\theta_0^* \in \Theta_0^*$, there is an initial condition $\hat s(0)$ such that $\|\hat s(0)\| \le \cdg{M_x}$ and the solution to \eqref{eq:bd} from $\hat s(0)$ satisfies
$$c_1 I_p \le \int_{\tau_0}^{\tau_0 + \tau} \psi^T(\hat s(t),t) \psi(\hat s(t),t) dt \le c_2 I_p, \quad \forall \tau_0 \ge 0.$$
\end{Assumption}	

Even if the assumption is formulated using a particular initial condition $\hat s(0)$, it will be seen in Section \ref{sec:3B} that $\psi$ is still persistently exciting with other $\hat s(0)$, thanks to Assumption \ref{ass:bd}.

One may be familiar with Assumption~\ref{ass:pe} from classical adaptive control.
However, unlike traditional adaptive control—where a fixed but unknown parameter \(\theta^*\) is estimated—here there is no fixed ground truth for \(\theta^*\). 
Instead, we should show that 
$$\theta_i(t) \quad \to \quad \cdg{\vartheta_\oo(t)} := \frac{1}{N} \sum_{i=1}^N \theta_i(t).$$
This implies that the target parameter \(\vartheta_\oo(t)\) is itself evolving, influenced by the individual variations of \(\theta_i(t)\). To address this challenge, we ensure that the adaptation process progresses gradually by restricting the adaptation gain \(g\) to be sufficiently small, aligning with the biological and sociological inspiration behind our approach. It is also a commonly accepted practice in adaptive control to have adaptation occur at a time scale slower than that of the system with no adaptation, this typically being achieved by requiring a small value of the adaptive gain. 
We show that this small value of $g$ also guarantees that $\vartheta_\oo(t)$ remains sufficiently close to $\theta_0^* = \vartheta_\oo(0)$ while all $\theta_i(t)$'s converge to each other.

Now, the next theorem ensures that the four assumptions introduced so far are enough for the proposed adaptation mechanism in the semi-global sense.

\begin{Theorem}\label{thm:main}
Suppose that Assumptions \ref{ass:graph}, \ref{ass:psi}, \ref{ass:bd}, and \ref{ass:pe} hold, and all the initial conditions satisfy $\|\theta_i(0)\| \le M_\theta$ and $\|x_i(0)\| \le M_x$ where \cdg{$M_\theta$} and $M_x$ are positive constants.
Define
\begin{multline*}
{\mathcal M} := \Big\{ (\theta_1,x_1,\cdots,\theta_N,x_N) : \\
\frac1N \sum_{i=1}^N \theta_i \in \Theta_0^*, \; \|\theta_i\| \le M_\theta, \; \|x_i\| \le M_x \Big\}.
\end{multline*}
Then, for any $\cdg{\delta} > 0$, there exists a threshold \(k^* \ge 1\) and a class-$\mathcal K$ function $\sigma$ such that, for \(k \ge k^*\) and \(g = 1/\sqrt{k}\), and for any initial condition in $\mathcal M$, the solutions \(x_i(t)\) and \(\theta_i(t)\) to \eqref{eq:mas} and \eqref{eq:adaptation_law} have the following properties:

\smallskip
\noindent 
(i) every state $x_i(t)$ converges to a neighborhood of $s(t)$ in the sense that 
\begin{equation}\label{eq:bigo}
\|x_i(t) - s(t)\| \le \frac{5\delta}{2}, \qquad \forall t \ge \sigma(1/k)
\end{equation}
(where $\sigma(1/k) \to 0$ as $k \to \infty$)

\smallskip
\noindent
(ii) every $\theta_i(t)$ converges to a constant, and $\vartheta_\oo(t)$ satisfies
\begin{equation}\label{eq:delta}
\|\vartheta_\oo(t) - \vartheta_\oo(0)\| \le \delta, \qquad \forall t \ge 0
\end{equation}

\smallskip
\noindent
(iii) it holds that
\begin{equation}\label{eq:claim12}
\lim_{t\to\infty} \|x_i(t) - s(t)\| = \lim_{t\to\infty} \|\theta_i(t) - \vartheta_\oo(t)\| = 0
\end{equation}
with an exponential convergence rate proportional to \(g\), for all $i=1,\cdots,N$,
\end{Theorem}

The proof of Theorem \ref{thm:main} is given in the next section, and we conclude this section with an intuitive motivation of the adaptation scheme \eqref{eq:adaptation_law}.
If all the states remain bounded, previous studies in \cite{kim2015a,lee2020a,lee2022b} reveal that there exists a threshold $k^*$ such that $\|x_i(t) - s(t)\| = O(1/k)$ and $\|\dot x_i(t) - \dot s(t)\| = O(1/k)$, $\forall i$, for $k \ge k^*$ and $t \ge \sigma(1/k)$. (The latter property was simply noted in \cite{lee2022b}, but can in fact be proved with some effort.)
This means that, with sufficiently large $k$, it holds that
$$\dot x_i = \psi(x_i,t) \theta_i + k \sum_{j \in \cN_i} (x_j-x_i) \approx \dot s = \psi(s,t) \vartheta_\oo$$
for $t \ge \sigma(1/k)$, which in turn implies
$$k \sum_{j \in \cN_i} (x_j-x_i) \approx \psi(s,t) \vartheta_\oo - \psi(x_i,t) \theta_i.$$
This means that the discrepancy between the vector field of the blended dynamics and that of each individual agent is compensated by the coupling terms.
Since $x_i \approx s$ for $t \ge \sigma(1/k)$, we then have
$$k \sum_{j \in \cN_i} (x_j-x_i) \approx \psi(s,t) (\vartheta_\oo - \theta_i).$$
Substituting this into \eqref{eq:adaptation_law} gives
$$\dot \theta_i \approx - g \psi^T(s,t) \psi(s,t) (\theta_i - \vartheta_\oo).$$
Therefore, if \(\vartheta_\oo\), the mean of the $\theta_i$, varies slowly, then the persistent excitation condition (Assumption \ref{ass:pe}) ensures the convergence of \(\theta_i(t)\) toward \(\vartheta_\oo(t)\).\footnote{The vector $\vartheta_\oo(t)$ need not lie in the set $\Theta_0^*$, but it remains close to $\Theta_0^*$. Hence, we can exploit the strict positivity of $c_1$.}
In summary, $x_i$ quickly approaches $s$, and since $\psi(s,t)$ itself is persistently exciting, $\psi(x_i,t)$ inherits this excitation, which in turn enables the slow adaptation of $\theta_i$, exhibiting a three-time-scale behavior.
This was our initial intuition in this study, but it lacked a rigorous analysis.
A formal analysis is outlined in the next section.

\section{Proof of Theorem 1}

\subsection{Analytical Framework}

In this subsection, we change the coordinates of the overall system so that the system appears in a convenient form for analysis, with the term $\psi^T(s,t) \psi(s,t)$ in Assumption~\ref{ass:pe} appearing explicitly.

By Assumption \ref{ass:graph}, the graph Laplacian matrix \(\mathcal{L}\) of \(\mathcal{G}\) admits a decomposition with a matrix \( \cdg{R} \in \R^{N \times (N-1)} \) such that \( R^T R = I_{N-1} \), \( R^T 1_N = 0 \) (where \( 1_N \) is the \(N\)-dimensional column vector of ones), and
\[
\begin{bmatrix} \frac{1}{N} 1_N^T \\ R^T \end{bmatrix} \mathcal{L} \begin{bmatrix} 1_N & R \end{bmatrix} =
\begin{bmatrix} 0 & 0 \\ 0 & \Lambda \end{bmatrix},
\]
with \cdg{\(\Lambda\)} being a positive definite diagonal matrix. 
Thus, \(\Lambda = R^T \mathcal{L} R\) and \(\mathcal{L} = R \Lambda R^T\).

To facilitate the analysis, we introduce the coordinate transformation:
\begin{subequations}\label{eq:def_chi}
\begin{equation}\label{eq:def_chia}
\begin{bmatrix} \cdg{\chi_\oo} \\ \cdg{\tilde \chi} \end{bmatrix} =
\begin{bmatrix} \frac{1}{N} 1_N^T \\ R^T \end{bmatrix}
\begin{bmatrix} x_1 \\ \vdots \\ x_N \end{bmatrix}
\quad \text{or} \quad
\begin{bmatrix} x_1 \\ \vdots \\ x_N \end{bmatrix} = 1_N \chi_\oo + R \tilde \chi.
\end{equation}
Hence, each \(x_i\) can be written as
\begin{equation}
x_i = \chi_\oo + r_i \tilde \chi,
\end{equation}
\end{subequations}
where \cdg{\(r_i\)} is the \(i\)-th row of \(R\).
Similarly, by defining \(\cdg{\theta} := \operatorname{col}(\theta_1,\dots,\theta_N)\), we introduce the transformed variables:\footnote{Recall that the Kronecker product \(\otimes\) satisfies \( (A \otimes B)(C \otimes D) = AC \otimes BD \), \(\|A \otimes B\| = \|A\| \|B\|\), and the eigenvalues of $A \otimes B$ are given by \(\{\mu_i \nu_j \}\) if the eigenvalues of \(A\) and \(B\) are \(\{\mu_i\}\) and \(\{\nu_j\}\), respectively.}
\begin{subequations}
\begin{equation}\label{eq:def_vartheta}
\begin{aligned}
\vartheta_\oo &= \frac{1}{N} (1_N^T \otimes I_p) \theta = \frac{1}{N} \sum_{i=1}^N \theta_i, & &\in \R^p, \\
\cdg{\tilde \vartheta} &= (R^T \otimes I_p) \theta, & &\in \R^{(N-1)p},
\end{aligned}
\end{equation}
so that
\begin{equation}\label{eq:def_vartheta_b}
\theta = 1_N \otimes \vartheta_\oo + (R \otimes I_p) \tilde\vartheta, \quad \text{or} \quad
\theta_i = \vartheta_\oo + (r_i \otimes I_p) \tilde\vartheta.
\end{equation}
\end{subequations}
Note importantly that \(\tilde \chi\) and \(\tilde \vartheta\) quantify the disagreement among the \(x_i\)'s and \(\theta_i\)'s, respectively. 
For example, if \(\tilde \vartheta = 0\), then all \(\theta_i\) are identical to \(\vartheta_\oo\), and \(\vartheta_\oo\) fully characterizes the synchronized behavior of $\theta_i$'s.

With this notation, and by defining \(\cdg{x} := \operatorname{col}(x_1,\dots,x_N)\), the system \eqref{eq:mas} can be rewritten as
\begin{align}\label{eq:tmp1}
\begin{split}
\dot x &= \begin{bmatrix} \psi(x_1,t) \theta_1 \\ \vdots \\ \psi(x_N,t) \theta_N \end{bmatrix} - k \mathcal{L} x \\
&= \left( I_N \otimes \psi(\chi_\oo,t) + \tilde\Psi_t(\chi_\oo,\tilde\chi) \right) \theta - k \mathcal{L} x,
\end{split}
\end{align}
where 
\begin{align}\label{eq:Psi}
\begin{split}
&\cdg{\tilde\Psi_t(\chi_\oo,\tilde\chi)} := \operatorname{diag} \{ \psi(\chi_\oo + r_1 \tilde\chi,t) - \psi(\chi_\oo,t), \\
&\qquad \dots, \psi(\chi_\oo + r_N \tilde\chi,t) - \psi(\chi_\oo,t) \} \quad \in \R^{N \times Np}.
\end{split}
\end{align}
Similarly, the adaptation law \eqref{eq:adaptation_law} becomes
\begin{align}\label{eq:tmp2}
\begin{split}
\dot \theta &= g \cdot {\rm diag}\{\psi^T(x_1,t), \cdots, \psi^T(x_N,t)\} \cdot (- k \mathcal{L} x) \\
	&= -g k \left( I_N \otimes \psi^T(\chi_\oo,t) + \tilde\Psi_t^T(\chi_\oo,\tilde\chi) \right) R \Lambda \tilde\chi,
\end{split}
\end{align}
where we have used \(\mathcal{L} = R\Lambda R^T\).

Using \eqref{eq:tmp1} and \eqref{eq:tmp2}\footnote{We also use the properties: $\frac1N 1_N^T (I_N \otimes \psi) \theta = (\frac1N 1_N^T \otimes 1) (I_N \otimes \psi) \theta = (1 \otimes \psi) (\frac1N 1_N^T \otimes I_p) \theta = \psi \vartheta_\oo$, $(R^T \otimes 1) (I_N \otimes \psi) \theta = (I_{N-1} \otimes \psi) (R^T \otimes I_p) \theta = (I_{N-1} \otimes \psi) \tilde\vartheta $, and $(1_N^T \otimes I_p)(I_N \otimes \psi^T) (R \otimes 1) = 0$.}, we derive the dynamics for \(\chi_\oo\), \(\tilde \chi\), \(\vartheta_\oo\), and \(\tilde\vartheta\):
\begin{subequations}\label{eq:sys}
\begin{align}
\dot \chi_\oo &= \psi(\chi_\oo,t) \vartheta_\oo + \frac{1}{N} 1_N^T \tilde\Psi_t(\chi_\oo,\tilde\chi) \theta, \label{eq:sys1} \\
\dot {\tilde\chi} &= - k \Lambda \tilde\chi + (I_{N-1} \otimes \psi(\chi_\oo,t)) \tilde\vartheta + R^T \tilde\Psi_t(\chi_\oo,\tilde\chi) \theta, \label{eq:sys2} \\
\dot \vartheta_\oo &= -\frac{gk}{N} (1_N^T \otimes I_p) \tilde \Psi_t^T(\chi_\oo,\tilde\chi) R \Lambda \tilde\chi, \label{eq:sys3} \\
\dot {\tilde \vartheta} &= - gk (I_{N-1} \otimes \psi^T(\chi_\oo,t)) \Lambda \tilde\chi \notag \\
	&\qquad -gk (R^T \otimes I_p) \tilde\Psi_t^T(\chi_\oo,\tilde\chi) R \Lambda \tilde\chi. \label{eq:sys4}
\end{align}
\end{subequations}
Note that \((\tilde\chi, \tilde\vartheta) = (0,0)\) is an equilibrium of the coupled system \eqref{eq:sys2} and \eqref{eq:sys4} (recall that \(\tilde\Psi_t(\chi_\oo,0)=0\)). This means that exact consensus of \(x_i\) and \(\theta_i\) is achievable. Provided consensus is reached, \(\vartheta_\oo\) remains constant due to \eqref{eq:sys3}, and the collective behavior is governed by the blended dynamics:
\[
\dot \chi_\oo = \psi(\chi_\oo,t) \vartheta_\oo,
\]
which is contained in \eqref{eq:sys1}. 
We see heuristically that the term \(- k \Lambda \tilde\chi\) in \eqref{eq:sys2} should push \(\tilde\chi\) towards zero because \(\Lambda\) is positive definite. 
The first term in \eqref{eq:sys1} yields the stability of \(\chi_\oo\) with Assumption~\ref{ass:bd}, provided that \(\vartheta_\oo(t)\) remains close to its initial value \(\vartheta_\oo(0)\). 
This last condition tends to be achieved when the adaptation gain \(g\) is small and \(\tilde\chi\) converges quickly to zero.

However, it is less immediate from \eqref{eq:sys4} that $\|\tilde\vartheta\|$ decreases, because unlike \(\tilde\chi\) (which benefits directly from the term \(-k\Lambda\tilde\chi\)), the reduction of $\|\tilde\vartheta\|$ depends on its coupling with \(\tilde\chi\).
Specifically, \(\tilde\vartheta\) appears in \eqref{eq:sys2} via the term \((I_{N-1} \otimes \psi(\chi_\oo,t)) \tilde\vartheta\), which then influences the evolution of \(\tilde\chi\). 
In light of the above heuristic argument, to elucidate this interaction we make a coordinate change from $\tilde\chi$ to \(\xi\) defined by:
\begin{equation}\label{eq:xi}
\xi := \tilde\chi - \frac{1}{k} (\Lambda^{-1} \otimes \psi(\chi_\oo,t)) \tilde\vartheta.
\end{equation}
With this transformation, the system \eqref{eq:sys} is rewritten as
\begin{subequations}\label{eq:syst}
\begin{align}
\dot \chi_\oo &= \psi(\chi_\oo,t) \vartheta_\oo + \frac{1}{N} 1_N^T \tilde\Psi_t(\chi_\oo,\tilde\chi) \theta, \label{eq:syst1} \\
\dot \vartheta_\oo &= - \frac{gk}{N} (1_N^T \otimes I_p) \tilde\Psi_t^T(\chi_\oo,\tilde\chi) R \Lambda \xi \notag \\
	&\quad - \frac{g}{N} (1_N^T \otimes I_p) \tilde\Psi_t^T(\chi_\oo,\tilde\chi) (R \otimes \psi(\chi_\oo,t)) \tilde\vartheta, \label{eq:syst2} \\
\dot {\tilde\vartheta} &= -gk (\Lambda\!\otimes\!\psi^T(\chi_\oo,\!t)) \xi - g (I_{N-1}\!\otimes\! {\psi^T(\chi_\oo,\!t)\psi(\chi_\oo,\!t)} ) \tilde\vartheta \notag \\
	&\quad -gk (R^T \otimes I_p) \tilde\Psi_t^T(\chi_\oo,\tilde\chi) R \Lambda \xi \notag \\
	&\quad - g (R^T \otimes I_p) \tilde\Psi_t^T(\chi_\oo,\tilde\chi) (R \otimes \psi(\chi_\oo,t)) \tilde\vartheta, \label{eq:syst3} \\
\dot \xi &= -k \Lambda \xi + R^T \tilde\Psi_t(\chi_\oo,\tilde\chi)\theta - \frac{1}{k} \left(\Lambda^{-1} \! \otimes \! \frac{d}{dt}\psi(\chi_\oo,t) \right) \! \tilde\vartheta \notag \\
	&\quad + g \|\psi(\chi_\oo,t)\|^2 \xi + \frac{g}{k}\|\psi(\chi_\oo,t)\|^2(\Lambda^{-1} \otimes \psi(\chi_\oo,t)) \tilde\vartheta \notag \\
	&\quad + g(\Lambda^{-1}R^T \otimes \psi(\chi_\oo,t)) \tilde\Psi_t^T(\chi_\oo,\tilde\chi) R\Lambda \xi \notag \\
	&\quad + \frac{g}{k}(\Lambda^{-1} R^T \otimes \psi(\chi_\oo,t)) \tilde\Psi_t^T(\chi_\oo,\tilde\chi) (R \otimes \psi(\chi_\oo,t)) \tilde\vartheta. \label{eq:syst4}
\end{align}
\end{subequations}
Note that, thanks to \eqref{eq:xi}, the term $\psi^T \psi$ appears in \eqref{eq:syst3}, which helps drive \(\tilde\vartheta\) to zero through the persistent excitation property of \(\psi\) (Assumption~\ref{ass:pe}).

\subsection{Uniform Exponential Stability from Persistent Excitation}\label{sec:3B}

Typically, when a regressor $\psi(t)$ is persistently exciting, it can be shown that a linear time-varying (LTV) system $\dot w = - \psi^T(t) \psi(t) w$ is exponentially stable \cite{anderson1977,moreno2021}.
In our case, persistent excitation of the regressor depends on a particular trajectory $\hat s(t)$ (see Assumption \ref{ass:pe}), and therefore, the corresponding LTV system also depends on the particular trajectory.
In this subsection, we consider an arbitrary solution $\s(t)$ to
\begin{equation}\label{eq:mys}
\dot \s = \psi(\s,t) \theta_0^*
\end{equation}
with \(\| \cdg{\s(0)} \| \le M_x\) and $\theta_0^* \in \Theta_0^*$, and we show that, leveraging Assumption~\ref{ass:bd}, the LTV systems corresponding to all such trajectories $\s(t)$ are all exponentially stable, and in addition, there is a family of Lyapunov functions with uniform lower/upper bounds and identical decay rate.

We first note uniform boundedness of any solution \(\s(t)\).
Indeed, writing
\begin{align}
\dot \s &= (\psi(\s,t)-\psi(0,t)) \theta_0^* + \psi(0,t) \theta_0^* \notag \\
&= \frac{\partial \psi}{\partial \s}(\check s,t) \theta_0^* \cdot \s + \psi(0,t) \theta_0^*, \label{eq:hats}
\end{align}
with some \(\check s\), we obtain, using \eqref{eq:c} and Assumption \ref{ass:psi},
\begin{equation}\label{eq:psi0}
\frac{d}{dt}\|\s\| = \frac{\s \dot \s}{\|\s\|} \le -c \|\s\| + \rho(0) M_\theta.
\end{equation}
Therefore, $\sup_{t \ge 0} \|\s(t)\| \le \cdg{M_{s}} := \max\{ M_x, \rho(0) M_\theta / c \}$.
Moreover, since $\|\hat s(0)\| \le M_x$, the solution \(\hat s(t)\) of \eqref{eq:bd} also satisfies \(\sup_{t \ge 0}\|\hat s(t)\| \le M_s\).
Define $\cdg{M_\psi} := \rho(M_s+\delta)$ (with the margin $\delta$ from Theorem \ref{thm:main}) so that $\|\psi(\s(t),t)\| \le M_\psi$, $\forall t$.

\begin{Lemma}\label{lem:2}
Suppose that Assumptions \ref{ass:bd} and \ref{ass:pe} hold, and let \(\s(t)\) be an arbitrary solution to \eqref{eq:mys} with \(\|\s(0)\| \le M_x\) and $\theta_0^* \in \Theta_0^*$.
Let \(\Phi_\s^g\) be the state-transition matrix of a linear time-varying system
\begin{equation}\label{eq:fs}
\dot w = \Big( - g \psi^T(\s(t),t) \psi(\s(t),t) \Big) \; w =: F_{\s}^g(t) \; w.
\end{equation}
Then there exist constants \(m, b > 0\) such that
\begin{equation}\label{eq:m}
\|\Phi_\s^g(t_1,t_0)\| \le m e^{- gb (t_1-t_0)}, \quad \forall t_1 \ge t_0 \ge 0,
\end{equation}
which holds for any $0 < g \le 1$ and any such solution $\s(t)$.
\end{Lemma}

The Lemma's proof can be found in the Appendix of \cite{appendix}.

Now, we define 
\begin{equation}\label{eq:ps}
\cdg{P_\s^g(t)} := g \int_t^\infty \Phi_\s^{gT}(\tau, t) \Phi_\s^g(\tau, t) \, d\tau
\end{equation}
for which, it holds that
\begin{equation}\label{eq:pdot}
\dot P_\s^g = - P_\s^g F_{\s}^g - F_{\s}^{gT} P_\s^g - g I_p.
\end{equation}
Moreover, we have
\begin{align*}
\|P_\s^g(t)\| &\le g \int_t^\infty m^2 e^{-2gb(\tau-t)} d\tau \le \frac{gm^2}{2gb} = \frac{m^2}{2b} =: \cdg{\lambda_M}, 
\end{align*}
and it can be shown (as in \cite[Theorem 4.12]{Khalil2002}) that
$$P_\s^g(t) \ge \frac{g}{2 g M_\psi^2} I_p = \frac{1}{2 M_\psi^2} I_p =: \cdg{\lambda_m} I_p$$
since $\|F_\s^g(t)\| \le g M_\psi^2$. 
Therefore,
\begin{equation}\label{eq:ulp}
\lambda_m I_p \le P_\s^g(t) \le \lambda_M I_p,
\end{equation}
for all \(t \ge 0\) and $0 < g \le 1$.
We emphasize that, although \(F_{\s}^g\) and \(P_\s^g\) depend on the initial condition \(\s(0)\) and the parameter $g$, the bounds \(\lambda_m\) and \(\lambda_M\) in \eqref{eq:ulp} are uniform for all initial conditions with \(\|\s(0)\| \le M_x\) and for all $0 < g \le 1$.

\subsection{Sketch of the remaining proof}\label{sec:III-C}

Due to the page limit, here we just outline the remaining part of the proof and refer the reader to the complete proof in \cite{appendix}.
In the complete proof, considerable effort is devoted to showing that all the variables $\chi_\oo$, $\tilde\chi$, $\vartheta_\oo$, $\tilde\vartheta$, $\xi$, and $s$ (the solution of \eqref{eq:bd0}) are uniformly bounded, and thus, $\exists L_\psi$ such that $\|\psi(\chi_\oo,t) - \psi(s,t)\| \le L_\psi \|\chi_\oo - s\|$.
In \cite{appendix}, we also show that $\sup_{t \ge 0}\|\vartheta_\oo(t) - \vartheta_\oo(0)\| \le \delta_\oo$ where $\delta_\oo \le \min\{c/(2L_\psi),\delta/2\}$.
Then we employ a Lyapunov function
$$V(\xi,\tilde\vartheta) = \frac{k^2}{2} \xi^T \xi + \frac{1}{2} \tilde\vartheta^T (I_{N-1}\otimes P_\s^g(t)) \tilde\vartheta$$
and show that $\dot V \le -g/(6 \lambda_M) V$ using \eqref{eq:pdot} and \eqref{eq:ulp}.
This means that \(\xi(t) \to 0\) and \(\tilde\vartheta(t) \to 0\) exponentially fast, with a rate proportional to \(g\); consequently, by \eqref{eq:xi}, \(\tilde\chi(t) \to 0\) exponentially fast. 
Moreover, one can show that, by adding and subtracting $(\psi(\chi_\oo,t) - \psi(s,t))\vartheta_\oo(0)$ to $\dot \chi_\oo - \dot s$,
\begin{align*}
\frac{d}{dt}\|\chi_\oo - s\| \le -c \|\chi_\oo - s\| + \delta_\oo L_\psi \|\chi_\oo - s\| + \frac{L_{\tilde\Psi} G_\theta \|\tilde\chi\|}{\sqrt{N}}
\end{align*}
where $G_\theta$ is an upper bound of $\|\theta(t)\|$, $L_{\tilde\Psi}$ comes from $\|\tilde\Psi_t(\chi_\oo,\tilde\chi)\| \le L_{\tilde\Psi} \|\tilde\chi\|$, and $c-\delta_\oo L_\Psi \ge c/2$.
Then, the difference between \(\chi_\oo(t)\) and \(s(t)\) converges exponentially to zero by the fact that the state of an exponentially stable system with exponentially decaying input goes to zero, and by the comparison lemma \cite{Khalil2002}.
Therefore, the claim (iii) follows from \(x_i(t) = \chi_\oo(t) + r_i \tilde\chi(t)\), and from \(\theta_i(t) = \vartheta_\oo(t) + (r_i \otimes I_p) \tilde\vartheta(t)\). 
Finally, from \eqref{eq:tmp2} it is apparent that \(\dot \theta(t) \to 0\) exponentially fast, proving that \(\theta(t)\) converges to a constant.

\section{Simulation Study}

\begin{figure}[t]
\centering
\includegraphics[width=0.36\textwidth]{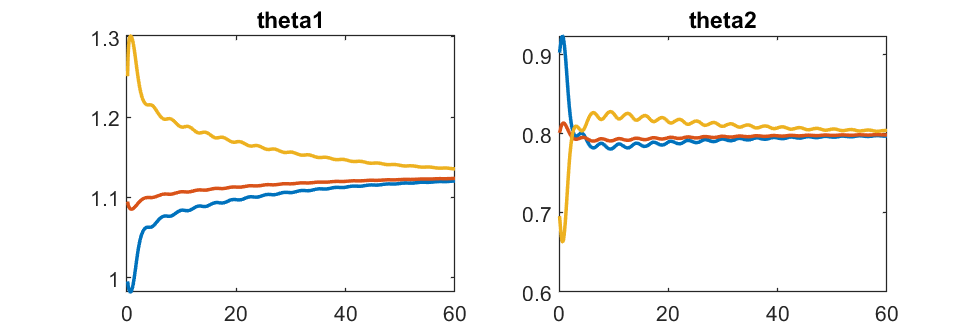}
\caption{Progress of two components of $\theta_i(t)$, $i=1,2,3$}
\label{fig:2}
\end{figure}

Consider a multi-agent system given by
$$\dot x_i = [-x_i, \; \sin(t)] \; \theta_i + k \textstyle{\sum_{j \in \cN_i}} (x_j - x_i)$$
where $\theta_i \in \R^2$.
Let $\Theta_0^*$ be any compact set in $\{ [\alpha,\beta]^T : \alpha \ge d_\alpha, |\beta| \ge d_\beta \}$ with arbitrary positive numbers $d_\alpha$ and $d_\beta$.
Then Assumption \ref{ass:bd} is verified.
Assumption \ref{ass:pe} also holds because \eqref{eq:bd} becomes $\dot {\hat s} = -\alpha \hat s + \beta \sin(t)$ and there is a particular solution $\hat s(t)$ satisfying Assumption \ref{ass:pe}.
Fig.~\ref{fig:2} shows convergence of parameters (under the ring graph when $N=3$), and more detailed simulations and codes are in \cite{github}.

Although only first-order node dynamics \eqref{eq:mas} are studied in this paper, additional simulation results in \cite{github} demonstrate that the proposed adaptation mechanism still works for higher-order systems when we interpret all the assumptions with $x_i \in \R^n$ and $\psi(x_i,t) \in \R^{n \times p}$ (Assumption \ref{ass:bd} in this case implies contractivity with a state-independent metric\footref{footnote1}). 

However, the next example motivates further study, and Assumption \ref{ass:bd} may need to be replaced with other stability conditions.
Consider coupled Van der Pol oscillators, described by 
$$\ddot x_i + \mu_i (x_i^2 - 1) \dot x_i + \nu_i x_i = k \sum_{j \in \cN_i} \big( x_j + \dot x_j - x_i - \dot x_i \big).$$
This can be written as, with \(z_i = x_i\) and \(y_i = x_i + \dot x_i\),
\begin{align*}
\dot z_i &= - z_i + y_i, \qquad\qquad i=1, \cdots, N, \\
\dot y_i &= (1-\mu_i(z_i^2-1)) (-z_i + y_i) - \nu_i z_i + k \sum_{j \in \cN_i} \big( y_j - y_i \big).
\end{align*}
Although each agent is second order, if we treat \(z_i\) as a function of $t$, the scalar dynamics of \(y_i\) can be written in the form of \eqref{eq:sys0} with
\begin{align*}
\psi(y_i,z_i(t)) &= \left[ -(z_i^2(t)-1)(-z_i(t) + y_i), -z_i(t) \right], \\
\theta_i &= [\mu_i, \nu_i]^T, \\
\psi_o(y_i,z_i(t)) &= -z_i(t) + y_i.
\end{align*}
So we treat $z_i(t)$ as a time function and the function $\psi(y_i,z_i(t))$ is treated as $\psi(y_i,t)$. 
However, the contractivity of Assumption \ref{ass:bd} does not hold because, on the limit cycle, two different solutions do not converge to each other.
Nevertheless, the simulation results are promising, as illustrated in Fig.~\ref{fig:1}.
This observation motivates further study, and stability of the limit cycle may play the role of Assumption \ref{ass:bd}, which needs more investigations.

\section{Conclusion}

Consensus has been studied from various perspectives in the literature, and it has been argued that enforced consensus in heterogeneous systems leads to emergent behavior \cite{lee2020a,lee2022b}. This paper demonstrates an additional role of consensus: enforced consensus can also serve as a mechanism for adaptation across heterogeneous agents, driving their individual dynamics toward a common vector field.


\newpage

{\small

\section*{Appendix: Proof of Lemma \ref{lem:2}}

The first step is to inspect the state-transition matrix $\Phi_{\hat s}^g$, which is for the LTV system
$$\dot w = F_{\hat s}^g(t) w = -g \psi^T(\hat s(t),t) \psi(\hat s(t),t) w$$
with $\hat s(t)$ of Assumption \ref{ass:pe}, based on the following lemma.

\medskip

\noindent{\bf Lemma A.1 \cite{moreno2021}}: {\it
The state-transition matrix $\Phi$ of the LTV system
$$\dot w = - \psi^T(t) \psi(t) \; w,$$
where $\psi(t)$ such that $\|\psi(t)\| \le M_\psi$, $\forall t$, and 
$$c_1 I_p \le \int_{\tau_0}^{\tau_0 + \tau} \psi^T(t) \psi(t) dt \le c_2 I_p, \qquad \forall \tau_0 \ge 0$$
in which, $\tau > 0$ and $0 < c_1 \le c_2$, satisfies 
$$\|\Phi(t_1,t_0)\| \le \sqrt{\frac{\kappa_1}{\kappa_2}} \exp\left( - \frac{c_1}{4\kappa_1} (t_1 - t_0) \right)$$
where
\begin{align*}
\kappa_1 &= \frac{2 (\tau M_\psi c_2)^2 + \tau c_1}{c_1} + \tau c_2, \qquad \text{and} \\
\kappa_2 &= \frac{2 (\tau M_\psi c_2)^2 + \tau c_1}{c_1}.
\end{align*}}

\medskip

Since Assumption \ref{ass:pe} implies
$$(g c_1) I_p \le \int_{\tau_0}^{\tau_0 + \tau} g\psi^T(\hat s(t),t)\psi(\hat s(t),t) dt \le (g c_2) I_p$$
we can employ Lemma A.1 with $c_1$, $c_2$, and $M_\psi$ replaced by $gc_1$, $gc_2$, and $\sqrt{g} M_\psi$, respectively.
This yields
\begin{equation}\label{eq:eq7}
\|\Phi_{\hat s}^g(t_1,t_0)\| \le \sqrt{\frac{\kappa_1^g}{\kappa_2^g}} \exp\left( - \frac{g c_1}{4\kappa_1^g} (t_1 - t_0) \right) \le \hat m e^{- g \hat b (t_1 - t_0)}
\end{equation}
where $\hat m$ and $\hat b$ are taken as follows.
Note that, for $0 < g \le 1$,
\begin{align*}
\kappa_1^g = \frac{2 g^3 (\tau M_\psi c_2)^2 + g \tau c_1}{g c_1} + g \tau c_2 \le \kappa_1
\end{align*}
so that 
$$\frac{c_1}{4\kappa_1^g} \ge \frac{c_1}{4\kappa_1} =: \hat b.$$
Also note that
\begin{align*}
\frac{\kappa_1^g}{\kappa_2^g} &= 1 + \frac{gc_1c_2}{2\tau g^2 M_\psi^2 c_2^2 + c_1} \\
&\le 1 + \frac{gc_1c_2}{2 \sqrt{2\tau c_1} g M_\psi c_2} = 1 + \frac{c_1}{2\sqrt{2\tau c_1} M_\psi} =: \hat m^2
\end{align*}
(in which we used $a^2 + b^2 \ge 2ab$).

Now, based on \eqref{eq:eq7}, we prove \eqref{eq:m} using the contraction property from Assumption \ref{ass:bd}.
Without loss of generality, we assume that $\hat b < c$ (otherwise, choose smaller $\hat b$).
Note that 
\[
\|\hat s(t) - \s(t)\| \le e^{-ct} \|\hat s(0) - \s(0)\| \le 2 M_x e^{-ct}, \quad \forall t\ge 0.
\]
Since a solution to
\[
\dot w = F_{\s}^g(t) w = F_{\hat s}^g(t) w + \left[F_{\s}^g(t) - F_{\hat s}^g(t)\right] w
\]
can be expressed in two different ways as
\begin{align*}
\Phi_{\s}^g(t_1,t_0) \bar w &= \Phi_{\hat s}^g(t_1,t_0) \bar w \\
&\; + \int_{t_0}^{t_1} \Phi_{\hat s}^g(t_1,\check t) \left[F_\s^g(\check t) - F_{\hat s}^g(\check t)\right] \Phi_\s^g(\check t,t_0) \bar w \, d{\check t}
\end{align*}
for any vector $\bar w$, it follows that
\begin{align*}
\|\Phi_{\s}^g(t_1,t_0)\| &\le \|\Phi_{\hat s}^g(t_1,t_0)\| \\
&\; + \int_{t_0}^{t_1} \|\Phi_{\hat s}^g(t_1,\check t)\| \|F_{\s}^g(\check t) - F_{\hat s}^g(\check t)\| \|\Phi_\s^g(\check t,t_0)\| \, d{\check t}.
\end{align*}
Recall that, with $\delta$ of Theorem \ref{thm:main}, $M_{\psi} = \rho(M_s+\delta)$. 
Now, let \cdg{\(L_\psi\)} be a Lipschitz constant for \(\psi(\s,t)\) for all \(\|\s\| \le M_s + \delta\) and \(t \ge 0\).
Then one can show that 
\begin{equation}\label{eq:fdiff}
\|F_{\s}^g(t) - F_{\hat s}^g(t)\| \le 2 g M_\psi L_\psi \|\s(t)-\hat s(t)\| \le 4 g M_\psi L_\psi M_x e^{-ct}.
\end{equation}
Moreover, since \(\|\Phi_\s^g(\check t,t_0)\| \le 1\) for all \(\check t \ge t_0 \ge 0\) due to the structure of \(F_{\s}^g(t)\), combining these estimates with \eqref{eq:eq7}, we obtain
\begin{align*}
\|\Phi_{\s}^g(t_1,t_0)\| &\le \hat m e^{- g \hat b (t_1-t_0)} \\
&\qquad + 4 g M_{\psi} L_\psi M_x \hat m e^{- g \hat b t_1} \int_{t_0}^{t_1} e^{g \hat b \check t} e^{-c \check t} d{\check t} \\
&\le \left( \hat m + \frac{4 g M_{\psi} L_\psi M_x \hat m}{c-g\hat b} \right) e^{- g \hat b (t_1 - t_0)} \\
&\le \left( \hat m + \frac{4 M_{\psi} L_\psi M_x \hat m}{c-\hat b} \right) e^{- g \hat b (t_1 - t_0)} \\
&=: m e^{-gb(t_1 - t_0)}
\end{align*}
where we used $e^{-c t_0} \le 1$ and $e^{(g\hat b - c)t_1} \ge 0$ for any $t_1, t_0 \ge 0$.
Thus, \eqref{eq:m} holds, and Lemma \ref{lem:2} is proved.

\section*{Appendix: Extended version of Section \ref{sec:III-C}}

While Section \ref{sec:III-C} presented a brief overview of the proof, we present the full proof in this section.
The proof proceeds with an arbitrary initial condition $(\theta(0),x(0))$ in the set ${\mathcal M}$, and we will take a threshold $k^*$.
The procedure of taking $k^*$ is independent of particular choice of $(\theta(0),x(0))$ because we will use their norm bounds rather than particular values of the solution $(\theta(t),x(t))$.
Therefore, $k^*$ is valid for all initial conditions in ${\mathcal M}$.

\medskip

Since \(\|\theta_i(0)\| \le M_\theta\) and \(\|x_i(0)\| \le M_x\) for all \(i\), we have, by \eqref{eq:def_chia} and \eqref{eq:def_vartheta},
\begin{gather}\label{eq:inicondition}
\begin{split}
\|\chi_\oo(0)\| \le M_x, \quad \|\tilde\chi(0)\| \le \sqrt{N} M_x, \\
\|\vartheta_\oo(0)\| \le M_\theta, \quad \|\tilde\vartheta(0)\| \le \sqrt{N} M_\theta.
\end{split}
\end{gather}

Now, consider the following conditions:
\begin{subequations}\label{eq:TA}
\begin{align}
\|\chi_\oo(t)\| &< M_{s} + \delta =: \cdg{G_{\chi_\oo}}, \label{eq:TA1} \\
\|\tilde\chi(t)\| &< \sqrt{N} M_x + \delta =: \cdg{G_{\tilde\chi}}, \label{eq:TA2} \\
\|\vartheta_\oo(t)\| &< M_\theta + \delta =: \cdg{G_{\vartheta_\oo}}, \label{eq:TA3} \\
\|\tilde\vartheta(t)\| &< \sqrt{\frac{\lambda_M}{\lambda_m}} \left( \sqrt{N} M_\theta + \frac{\delta}{2}\right) + \delta =: \cdg{G_{\tilde\vartheta}}, \label{eq:TA4} \\
\|\vartheta_\oo(t) &- \vartheta_\oo(0)\| < \delta_\oo, \label{eq:TA5}
\end{align}
\end{subequations}
where 
\begin{equation}\label{eq:deltao}
\cdg{\delta_\oo} := \min \left\{ \delta, \;\; \frac{c}{2M_\psi}\delta, \;\; \frac{c}{16 \lambda_M M_\psi^2 L_{\psi}}, \frac{c}{2L_\psi} \right\}.
\end{equation}
Since the state trajectories are continuous, these conditions hold for a short time at least after \(t=0\). Let \(\cdg{T} > 0\) be the first time at which any one of \eqref{eq:TA} is violated (i.e., one of the inequalities becomes an equality), and assume that \(T\) is finite. In the following, we will show that \textit{the conditions in \eqref{eq:TA} continue to hold with strict inequality at \(t=T\) (hence, \(T = \infty\) by contradiction)}, provided \(k \ge k^*\) and \(g = 1/\sqrt{k}\), where \(k^*\) is defined as \cdr{\(k^* = \max\{1,k_1^*,\cdots,k_{10}^*\}\)} (each \(k_i^*\) is defined in the subsequent analysis).

For convenience, we introduce several definitions. By Assumption \ref{ass:psi} and \eqref{eq:TA}, there exist constants \cdg{\(M_{\tilde\Psi}\)} and \cdg{\(L_{\tilde\Psi}\)} such that 
\begin{align*}
\|\tilde\Psi_t(\chi_\oo,\tilde\chi)\| &\le M_{\tilde\Psi}, \\
\|\psi(\chi_\oo + r_i \tilde\chi,t) - \psi(\chi_\oo,t)\| &\le L_{\tilde\Psi} \|\tilde\chi\|, \quad \forall i.
\end{align*}
Thus, we have two bounds for $\tilde\Psi_t$:
\begin{equation}\label{eq:normPsi}
\|\tilde\Psi_t(\chi_\oo,\tilde\chi)\| \le M_{\tilde\Psi}, \qquad \|\tilde\Psi_t(\chi_\oo,\tilde\chi)\| \le L_{\tilde\Psi} \|\tilde\chi\| .
\end{equation}
We also note that, from \eqref{eq:syst1} and \eqref{eq:TA},
\begin{equation}\label{eq:C1}
\|\dot\chi_\oo\| \le M_{\psi} G_{\vartheta_\oo} + \frac{1}{\sqrt{N}} M_{\tilde\Psi} G_\theta =: \cdg{C_1},
\end{equation}
where \(\cdg{G_\theta} := \sqrt{N} G_{\vartheta_\oo} + G_{\tilde\vartheta}\) so that \(\|\theta(t)\| \le G_\theta\) by \eqref{eq:def_vartheta_b}.

From now on, we prove \eqref{eq:TA} in two different time intervals. In the first interval, both $\vartheta_\oo$ and $\tilde\vartheta$ remain near their initial values because the interval is sufficiently short (even if the gain $k$ is large in \eqref{eq:tmp2} or in \eqref{eq:sys3}--\eqref{eq:sys4}). The state $\chi_\oo$ also stays near its initial value, but the state $\tilde\chi$ decreases quickly so that $\|\tilde\chi(t)\|$ can be made sufficiently small by choosing a large enough $k$ at the end of this interval. In the second interval, we exploit the smallness of $\|\tilde\chi\|$ and show the convergence of $\xi$ and $\tilde\vartheta$ to zero using a Lyapunov function.

For simplicity, let us use \(\|f\|_{[a,b]} := \sup_{a \le t \le b} \|f(t)\|\). 
Let \cdg{\(\lambda_2\)} and \cdg{\(\lambda_N\)} denote the smallest and largest eigenvalues of \(\Lambda\), respectively, and define
\begin{equation}\label{eq:T0}
\cdg{T_0} = \frac{\delta_\oo/2}{(M_\psi + M_{\tilde\Psi}) \lambda_N G_{\tilde\chi}}.
\end{equation}

\subsection{For the interval \mathbookmark{$0 \le t \le T_0/\sqrt{k}$}{(eq)}}

In order, we establish \eqref{eq:TA2}, \eqref{eq:TA3}, \eqref{eq:TA4}, \eqref{eq:TA5}, and \eqref{eq:TA1}.

From \eqref{eq:sys2} and using \(\|R\| = 1\), we have
\[
\frac{d\|\tilde\chi\|}{dt} \le - k \lambda_2 \|\tilde\chi\| + M_\psi G_{\tilde\vartheta} + M_{\tilde\Psi} G_\theta,
\]
which implies
\begin{align}
\|\tilde\chi(t)\| &\le e^{-k \lambda_2 t} \|\tilde\chi(0)\| + \frac{M_\psi G_{\tilde\vartheta} + M_{\tilde\Psi} G_\theta}{k \lambda_2} (1 - e^{-k \lambda_2 t}) \notag \\
&\le e^{-k \lambda_2 t} \sqrt{N} M_x + \frac{C_2}{k} (1 - e^{-k \lambda_2 t}) \label{eq:tildechi} \\
&\le \max \left\{ \sqrt{N} M_x, \frac{C_2}{k} \right\} \le \sqrt{N} M_x < G_{\tilde\chi}, \label{eq:tildechi2}
\end{align}
for \(k \ge k^* \ge \cdr{k_1^*} = C_2/(\sqrt{N}M_x)\) (with \(\cdg{C_2}\) defined appropriately). 
Thus, \cdr{\eqref{eq:TA2} holds}. 

Next, from \eqref{eq:tmp2} we have
\[
\|\dot\theta\| \le \sqrt{k} (M_\psi + M_{\tilde\Psi}) \lambda_N G_{\tilde\chi},
\]
which yields, with \eqref{eq:T0},
$$\|\theta(t)-\theta(0)\|_{[0,T_0/\sqrt{k}]} \le \frac{T_0}{\sqrt{k}} \|\dot\theta\|_{[0,T_0/\sqrt{k}]} \le \delta_\oo/2.$$
(Here we can infer the motivation for the term $\sqrt{k}$ in the length $T_0/\sqrt{k}$ of the time interval.)
This further implies that
\begin{equation}\label{eq:vartheta}
\|\vartheta_\oo(t)-\vartheta_\oo(0)\|_{[0,T_0/\sqrt{k}]} \le \delta_\oo/2,
\end{equation}
and consequently,
\begin{align}
\|\vartheta_\oo\|_{[0,T_0/\sqrt{k}]} &\le M_\theta + \delta/2 < G_{\vartheta_\oo}, \\
\|\tilde\vartheta\|_{[0,T_0/\sqrt{k}]} &\le \sqrt{N} M_\theta + \delta/2 < G_{\tilde\vartheta}, \label{eq:31}
\end{align}
so that \cdr{\eqref{eq:TA3}, \eqref{eq:TA4}, and \eqref{eq:TA5} hold}.

Finally, to establish \eqref{eq:TA1}, introduce a reference signal \(\s(t)\) as the solution of \eqref{eq:mys} with initial condition \(\s(0) = \chi_\oo(0)\). 
Note that \(\theta_0^* = \vartheta_\oo(0) \in \Theta_0^*\).
Then we have using \eqref{eq:syst1}
\begin{align}
&\dot \chi_\oo - \dot \s = \psi(\chi_\oo,t)\vartheta_\oo(t) \! - \! \psi(\s,t)\vartheta_\oo(0) \! + \! \frac{1}{N} 1_N^T\tilde\Psi_t(\chi_\oo,\tilde\chi)\theta(t) \notag \\
&= \big( \psi(\chi_\oo,t) - \psi(\s,t) \big) \vartheta_\oo(0) + \psi(\chi_\oo,t) \big( \vartheta_\oo(t) - \vartheta_\oo(0) \big) \notag \\
&\qquad\qquad\qquad + \frac{1}{N} 1_N^T\tilde\Psi_t(\chi_\oo,\tilde\chi)\theta(t). \label{eq:xos}
\end{align}
Using \eqref{eq:TA5}, one can derive (using for the first summand \eqref{eq:c} in a manner similar to \eqref{eq:hats}) that
\begin{equation}\label{eq:chios}
\frac{d}{dt}\|\chi_\oo - \s\| \le -c \|\chi_\oo - \s\| + M_\psi \delta_\oo + \frac{1}{\sqrt{N}} L_{\tilde\Psi} \|\tilde\chi\| G_\theta.
\end{equation}
Then, by \eqref{eq:tildechi2} and the fact that \(\s(0)=\chi_\oo(0)\),
\begin{align}
\|\chi_\oo - \s\|_{[0,T_0/\sqrt{k}]} &\le \frac{T_0}{\sqrt{k}} \left( M_\psi \delta_\oo + L_{\tilde\Psi} M_x G_\theta \right) \label{eq:chios0} \\
&\le \delta/2,
\end{align}
with an appropriate choice of \(\cdr{k_2^*}\) so that the second inequality holds for $k \ge k_2^*$. Thus, \cdr{\eqref{eq:TA1} holds} since \(\|\s(t)\| \le M_s\) for all \(t \ge 0\).

The analysis so far shows that \(T > T_0/\sqrt{k}\) for \(k \ge k^*\).

\subsection{For the interval \mathbookmark{$T_0/\sqrt{k} \le t \le T$}{(eq)}}

In order, we will establish \eqref{eq:TA2}, \eqref{eq:TA1}, \eqref{eq:TA4}, \eqref{eq:TA5}, and \eqref{eq:TA3}.

From \eqref{eq:tildechi}, choose \(\cdr{k_3^*}\) large enough so that
\[
e^{-k \lambda_2 \frac{T_0}{\sqrt{k}}} \sqrt{N} M_{x} \le \frac{C_2}{k} < \frac{\delta}{2}, \quad \forall k \ge k_3^*.
\]
Then from \eqref{eq:tildechi} it follows that
\begin{equation}\label{eq:fastconv1}
\|\tilde\chi(t)\|_{[T_0/\sqrt{k},T]} \le \frac{2C_2}{k} < \delta,
\end{equation}
so that \cdr{\eqref{eq:TA2} holds at \(t=T\)}. This inequality, together with \eqref{eq:chios}, implies that for \(T_0/\sqrt{k} \le t \le T\),
\begin{align*}
\|\chi_\oo(t) - \s(t)\| &\le e^{-c(t-T_0/\sqrt{k})} \|\chi_\oo(T_0/\sqrt{k}) - \s(T_0/\sqrt{k})\| \\
&\quad + \left(\frac{M_\psi \delta_\oo}{c} + \frac{2 L_{\tilde\Psi} C_2 G_\theta}{k c \sqrt{N}}\right) (1 - e^{-c(t-T_0/\sqrt{k})}),
\end{align*}
which guarantees (using \eqref{eq:chios0}) that
\begin{align}
\|\chi_\oo(t) - \s(t)\|_{[T_0/\sqrt{k},T]} &\le \max \left\{ \frac{T_0}{\sqrt{k}} \left( M_\psi \delta_\oo + L_{\tilde\Psi} M_x G_\theta \! \right), \! \right. \notag \\
&\quad \left. \frac{M_\psi \delta_\oo}{c} + \frac{2 L_{\tilde\Psi} C_2 G_\theta}{k c \sqrt{N}} \right\} \notag \\
&< \min\left\{ \frac{1}{8 \lambda_M M_\psi L_{\psi}}, \delta \right\}, \label{eq:diffbound}
\end{align}
with a suitable choice of \(\cdr{k_4^*}\), recalling that the definition of \(\delta_\oo\) ensures
\[
\frac{M_\psi \delta_\oo}{c} \le \min \left\{ \frac{1}{16 \lambda_M M_\psi L_{\psi}}, \frac{\delta}{2} \right\}.
\]
Thus, \cdr{\eqref{eq:TA1} holds at \(t=T\)} (as ensured by \eqref{eq:diffbound}).

Next, we estimate the derivative of \(\|\xi(t)\|\) from \eqref{eq:syst4}. As a preliminary, we see using \eqref{eq:TA}, \eqref{eq:C1}, and Assumption \ref{ass:psi}, that there exists \cdr{\(M_{d\psi}\)} such that
\[
\left\| \! \frac{d}{dt}\psi^T \! (\chi_\oo,t) \right\| \le \left\| \frac{\partial \psi^T}{\partial s} \! (\chi_\oo,t) \right\| \! \| \dot \chi_\oo \| + \left\| \! \frac{\partial \psi^T}{\partial t} \! (\chi_\oo,t) \right\| \! \le \! M_{d\psi}.
\]
Also, using \eqref{eq:xi} and \eqref{eq:normPsi}, one obtains
\[
\|R^T\tilde\Psi_t(\chi_\oo,\tilde\chi)\theta\| \le L_{\tilde\Psi} \|\xi\| G_\theta + \frac{L_{\tilde\Psi} M_\psi \|\tilde\vartheta\| G_\theta}{k\lambda_2}.
\]
Then, from \eqref{eq:syst4} one can show that
\begin{equation}\label{eq:dotxi}
\xi^T\dot\xi \le - \left( k \lambda_2 - C_3 - g C_4 \right) \|\xi\|^2 + \left(\frac{C_5}{k} + \frac{g C_6}{k} \right)\|\xi\|\|\tilde\vartheta\|,
\end{equation}
where
\begin{gather*}
\cdg{C_3} = L_{\tilde\Psi} G_\theta, \quad \cdg{C_4} = M_\psi^2 + \frac{M_\psi M_{\tilde\Psi} \lambda_N}{\lambda_2}, \\
\cdg{C_5} = \frac{L_{\tilde\Psi} M_\psi G_\theta + M_{d\psi}}{\lambda_2}, \quad \cdg{C_6} = \frac{M_\psi^3 + M_\psi^2 M_{\tilde\Psi}}{\lambda_2}.
\end{gather*}
Inequality \eqref{eq:dotxi} implies that, with \(g = 1/\sqrt{k} \le 1\),
\[
\frac{d}{dt}\|\xi\| \le - \left( k \lambda_2 - C_7 \right) \|\xi\| + \frac{C_8}{k},
\]
where \(\cdg{C_7} = C_3 + C_4\) and \(\cdg{C_8} = (C_5 + C_6) G_{\tilde\vartheta}\). In addition, since it follows from \eqref{eq:xi} and $k \ge 1$ that
\[
\|\xi(0)\| \le \sqrt{N}M_x + \frac{M_\psi}{\lambda_2} \sqrt{N} M_\theta =: \cdg{M_\xi},
\]
solving the above differential inequality for \(0 \le t \le T_0/\sqrt{k}\) yields
\begin{align}\label{eq:C7}
\|\xi(T_0/\sqrt{k})\| &\le e^{-(k\lambda_2 - C_7) \frac{T_0}{\sqrt{k}}} M_\xi + \frac{C_8}{k(k\lambda_2 - C_7)} \notag \\
&\le \frac{2C_8}{k(k\lambda_2 - C_7)},
\end{align}
with an appropriate choice of \(\cdr{k_5^*}\).

Now let us proceed with \eqref{eq:syst3}.
First, define $F_{\chi_\oo}^g(t) = - g \psi^T(\chi_\oo(t),t) \psi(\chi_\oo(t),t)$.
Then we have 
\begin{align*}
\lambda_M \|I_{N-1} \! \otimes \! (F_{\chi_\oo}^g(t) \!- \! F_\s^g(t))\| &\le \lambda_M \| F_{\chi_\oo}^g(t) \!- \! F_\s^g(t) \| \\
\le 2 g \lambda_M M_\psi L_\psi \|\chi_\oo(t) \!-\! \s(t)\| &\le \frac{g}{4},
\end{align*}
which is derived similarly to the first inequality of \eqref{eq:fdiff} using \eqref{eq:diffbound}.
With \(F_\s^g\) in \eqref{eq:fs} and \(P_\s^g\) of \eqref{eq:ps}, one can show that, using \eqref{eq:ulp} and \eqref{eq:pdot} for the solution \(\s\) (such that $\s(0) = \chi_\oo(0)$)
\begin{align}
&\frac{1}{2} \tilde\vartheta^T (I_{N-1} \otimes \dot P_\s^g) \tilde\vartheta + \tilde\vartheta^T (I_{N-1} \otimes P_\s^g) \dot{\tilde\vartheta} \notag \\
&\le -\frac{g}{2} \|\tilde\vartheta\|^2 + \lambda_M \| I_{N-1} \otimes (F_{\chi_\oo}^g(t) - F_\s^g(t)) \| \|\tilde\vartheta\|^2 \notag \\
&\quad + gk C_9 \|\xi\|\|\tilde\vartheta\| + g \lambda_M M_\psi L_{\tilde\Psi} \left( \|\xi\| + \frac{M_\psi}{k\lambda_2} \|\tilde\vartheta\| \right) \|\tilde\vartheta\|^2 \notag \\
&\le -g \left( \frac{1}{4} - \frac{C_{10}}{k} \right) \|\tilde\vartheta\|^2 + \left( gk C_9 + g C_{11} \right) \|\xi\|\|\tilde\vartheta\|, \label{eq:dotvartheta}
\end{align}
where
\begin{gather*}
\cdg{C_9} = \lambda_M \lambda_N (M_\psi + M_{\tilde\Psi}), \quad \cdg{C_{10}} = \frac{\lambda_M M_\psi^2 L_{\tilde\Psi} G_{\tilde\vartheta}}{\lambda_2}, \\
\cdg{C_{11}} = \lambda_M M_\psi L_{\tilde\Psi} G_{\tilde\vartheta} .
\end{gather*}

Define the Lyapunov function
\begin{equation}\label{eq:V}
V(\xi,\tilde\vartheta) = \frac{k^2}{2} \xi^T \xi + \frac{1}{2} \tilde\vartheta^T (I_{N-1}\otimes P_\s^g(t)) \tilde\vartheta.
\end{equation}
Let
\[
\cdr{k_6^*} = \max \left\{ \frac{3 C_7}{\lambda_2}, 12 C_{10} \right\},
\]
so that \(k\lambda_2 - C_3 - g C_4 \ge k\lambda_2 - C_7 \ge \frac{2}{3} k\lambda_2\) and \(\frac{1}{4} - \frac{C_{10}}{k} \ge \frac{1}{6}\). 
With these, we obtain from \eqref{eq:dotxi} and \eqref{eq:dotvartheta}
\begin{align}
\dot V &\le - \frac{k^3 \lambda_2}{3} \|\xi\|^2 - \frac{g}{12} \|\tilde\vartheta\|^2 - \Bigg\{ \frac{k^3 \lambda_2}{3} \|\xi\|^2 + \frac{g}{12} \|\tilde\vartheta\|^2 \notag \\
&\qquad - \left( k C_5 + gk(C_6 + C_9) + g C_{11} \right) \|\xi\|\|\tilde\vartheta\| \Bigg\}. \label{eq:dotV}
\end{align}
Using \(g = 1/\sqrt{k}\), the expression in braces becomes nonnegative provided that
\[
\frac{\lambda_2}{36} k^{2.5} \ge \frac{1}{4}\left( k C_5 + \sqrt{k} (C_6+C_9) + \frac{C_{11}}{\sqrt{k}} \right)^2
\]
for all \(k \ge k_7^*\) with an appropriate choice of $\cdr{k_7^*}$.
Furthermore, with \(\cdr{k_8^*}\) chosen so that
\begin{equation}
\frac{2\lambda_2 k}{3} \ge \frac{g}{6\lambda_M} = \frac{1}{6\lambda_M \sqrt{k}}, \quad \forall k \ge k_8^*,
\end{equation}
inequality \eqref{eq:dotV} yields
\begin{align}\label{eq:decV}
\begin{split}
\dot V &\le - \frac{2\lambda_2 k}{3} \cdot \frac{k^2}{2} \|\xi\|^2 - \frac{g}{6\lambda_M} \cdot \frac{\lambda_M}{2} \|\tilde\vartheta\|^2 \\
&\le - \frac{g}{6\lambda_M} \left( \frac{k^2}{2} \|\xi\|^2 + \frac{\lambda_M}{2} \|\tilde\vartheta\|^2 \right) \le -\frac{g}{6 \lambda_M} V.
\end{split}
\end{align}
Thus, for \(T_0/\sqrt{k} \le t \le T\),
\begin{align*}
\frac{\lambda_m}{2} \|\tilde\vartheta(t)\|^2 &\le V(t) \le V(T_0/\sqrt{k}) \\
&\le \frac{k^2}{2} \|\xi(T_0/\sqrt{k})\|^2 + \frac{\lambda_M}{2} \|\tilde\vartheta(T_0/\sqrt{k})\|^2.
\end{align*}
Now take \(\cdr{k_9^*}\) such that
\[
\frac{2 C_8}{(k_9^* \lambda_2-C_7)\sqrt{\lambda_m}} < \delta.
\]
Then, by \eqref{eq:C7}, and by the left inequality of \eqref{eq:31} at $t = T_0/\sqrt{k}$, the above inequality yields
\[
\|\tilde\vartheta(t)\| \le \frac{2 C_8}{(k\lambda_2-C_7)\sqrt{\lambda_m}} + \sqrt{\frac{\lambda_M}{\lambda_m}} \left(\sqrt{N} M_\theta + \frac{\delta}{2}\right) < G_{\tilde\vartheta}.
\]
Hence, \cdr{\eqref{eq:TA4} holds at \(t=T\)}.

Next, defining
\[
\cdg{C_{12}} = \frac{2L_{\tilde\Psi}}{\sqrt{N}} \left( \lambda_N + \frac{\lambda_N M_\psi}{\lambda_2 \sqrt{\lambda_m}} + \frac{M_\psi}{\sqrt{\lambda_m}} + \frac{M_\psi^2}{\lambda_2 \lambda_m} \right),
\]
observe the following:
\begin{align}
&\|\vartheta_\oo(t) - \vartheta_\oo(T_0/\sqrt{k})\|_{[T_0/\sqrt{k},T]} \notag \\
&\underset{\eqref{eq:syst2}}{\le} \! \int_{\frac{T_0}{\sqrt{k}}}^T \! \left\| \frac{g}{N}(1_N^T \! \otimes \! I_p) \tilde\Psi_t^T(\chi_\oo, \! \tilde\chi) \big(kR\Lambda\xi \! + \! (R \! \otimes \! \psi(\chi_\oo,t)\tilde\vartheta\big) \! \right\| \! dt \notag \\
&\underset{\eqref{eq:normPsi},\eqref{eq:xi}}{\le} \int_{\frac{T_0}{\sqrt{k}}}^T g \frac{L_{\tilde\Psi}}{\sqrt{N}}\left( \|\xi\| + \frac{M_\psi}{k\lambda_2} \|\tilde\vartheta\| \right) 
\left( k\lambda_N \|\xi\| + M_\psi \|\tilde\vartheta\| \right) dt \notag \\
&\underset{\eqref{eq:V}}{\le} \int_{\frac{T_0}{\sqrt{k}}}^T g \frac{L_{\tilde\Psi}}{\sqrt{N}}\left( \sqrt{\frac{2V}{k^2}} + \frac{M_\psi}{k\lambda_2} \sqrt{\frac{2V}{\lambda_m}} \right) \notag \\
&\qquad\qquad\qquad\qquad \times \left( k\lambda_N \sqrt{\frac{2V}{k^2}} + M_\psi \sqrt{\frac{2V}{\lambda_m}} \right) dt \notag \\
&= \frac{g}{k} C_{12} \int_{\frac{T_0}{\sqrt{k}}}^T V(t) dt \underset{\eqref{eq:decV}}{\le} \frac{g}{k} C_{12} V(T_0/\sqrt{k}) \int_{\frac{T_0}{\sqrt{k}}}^T e^{- \frac{g}{6\lambda_M} t} dt \notag \\
&\underset{\eqref{eq:V}}{\le} \frac{6\lambda_M}{k} C_{12} \left( \frac{k^2}{2} \|\xi(T_0/\sqrt{k})\|^2 + \frac{\lambda_M}{2} \|\tilde\vartheta(T_0/\sqrt{k})\|^2 \right) \notag \\
&= 3\lambda_M C_{12} k\|\xi(T_0/\sqrt{k})\|^2 + \frac{3\lambda_M^2 C_{12}}{k} \|\tilde\vartheta(T_0/\sqrt{k})\|^2 \notag \\
&\underset{\eqref{eq:C7}, \eqref{eq:31}}{\le} \frac{12\lambda_M C_{12} C_8^2}{k(k\lambda_2-C_7)^2} + \frac{3\lambda_M^2 C_{12} (\sqrt{N} M_\theta + \delta/2)^2}{k} \notag \\
&< \delta_\oo/2 \label{eq:hopelast}
\end{align}
with a choice of \(\cdr{k_{10}^*}\) guaranteeing the last inequality for $k \ge k_{10}^*$.
Thus, with \eqref{eq:vartheta}, \cdr{\eqref{eq:TA5} holds at \(t=T\)}, which also ensures that \cdr{\eqref{eq:TA3} holds at \(t=T\)}.

\bigskip

Therefore, by contradiction, we deduce that \(T=\infty\). 

This fact, with \eqref{eq:vartheta} and \eqref{eq:hopelast}, implies
$$\|\vartheta_\oo(t) - \vartheta_\oo(0)\| \le \delta_\oo \le \delta, \quad \forall t \ge 0.$$
Therefore, the second statement of Theorem \ref{thm:main}.(ii) is proved.
 
Now, adding and subtracting $\psi(s,t) \vartheta_\oo(0)$ to $\dot s - \dot \s$, we obtain that, by $\delta_\oo \le (c\delta)/(2M_\psi)$,
\begin{align*}
\frac{d}{dt} \|s - \s\| &\le -c \|s - \s\| + M_\psi \|\vartheta_\oo(t) - \vartheta_\oo(0)\| \\
&\le -c \|s - \s\| + c\frac{\delta}{2} \qquad \text{with $s(0) - \s(0) = 0$.}
\end{align*}
Therefore, $\|s(t)-\s(t)\| \le \delta/2$ for $t \ge 0$.
This fact, with \eqref{eq:fastconv1}, \eqref{eq:diffbound}, and $x_i = \chi_\oo + r_i \tilde\chi$, implies
\begin{align*}
\|x_i(t) - s(t)\| &\le \|\chi_\oo(t) - \s(t)\| + \|\s(t)-s(t)\| + \|\tilde\chi(t)\| \\
&\le \delta + \delta/2 + \delta
\end{align*}
for $t \ge T_0/\sqrt{k}$.
Therefore, Theorem \ref{thm:main}.(i) is proved.

From \eqref{eq:decV} together with \eqref{eq:V} and \eqref{eq:ulp}, it follows that \(\xi(t) \to 0\) and \(\tilde\vartheta(t) \to 0\) exponentially fast (with a rate proportional to \(g\)); consequently, by \eqref{eq:xi}, \(\tilde\chi(t) \to 0\) exponentially fast. 
Moreover, one can show that, by adding and subtracting $(\psi(\chi_\oo,t) - \psi(s,t))\vartheta_\oo(0)$ to $\dot \chi_\oo - \dot s$,
\begin{align*}
\frac{d}{dt}\|\chi_\oo - s\| &\le -c \|\chi_\oo - s\| + L_\psi \delta_\oo \|\chi_\oo - s\| + \frac{1}{\sqrt{N}} L_{\tilde\Psi} \|\tilde\chi\| G_\theta \\
&\le -\frac{c}{2} \|\chi_\oo - s\| + \frac{L_{\tilde\Psi}G_\theta}{\sqrt{N}} \|\tilde\chi\| .
\end{align*}
Then, the difference between \(\chi_\oo(t)\) and \(s(t)\) converges exponentially to zero by the fact that the state of an exponentially stable system with exponentially decaying input goes to zero, and by the comparison lemma \cite{Khalil2002}.
Theorem \ref{thm:main}.(iii) follows from \(x_i(t) = \chi_\oo(t) + r_i \tilde\chi(t)\), and from \(\theta_i(t) = \vartheta_\oo(t) + (r_i \otimes I_p) \tilde\vartheta(t)\). 

Finally, from \eqref{eq:tmp2} it is apparent that \(\dot \theta(t) \to 0\) exponentially fast, proving that \(\theta(t)\) converges to a constant (the first statement of Theorem \ref{thm:main}.(ii)). 
This completes the proof.


\begin{thebibliography}{10}

\bibitem{wieland2011}
P.~Wieland, R.~Sepulchre, and F.~Allg{\"o}wer, ``An internal model principle is
  necessary and sufficient for linear output synchronization,'' {\em
  Automatica}, vol.~47, no.~5, pp.~1068--1074, 2011.

\bibitem{kim2011}
H.~Kim, H.~Shim, and J.~H. Seo, ``Output {{Consensus}} of {{Heterogeneous
  Uncertain Linear Multi-Agent Systems}},'' {\em IEEE Transactions on Automatic
  Control}, vol.~56, pp.~200--206, Jan. 2011.

\bibitem{kim2015a}
J.~Kim, J.~Yang, H.~Shim, J.-S. Kim, and J.~H. Seo, ``Robustness of
  synchronization of heterogeneous agents by strong coupling and a large number
  of agents,'' {\em IEEE Transactions on Automatic Control}, vol.~61, no.~10,
  pp.~3096--3102, 2016.

\bibitem{panteley2017}
E.~Panteley and A.~Lor{\'i}a, ``Synchronization and {{Dynamic Consensus}} of
  {{Heterogeneous Networked Systems}},'' {\em IEEE Transactions on Automatic
  Control}, vol.~62, pp.~3758--3773, Aug. 2017.

\bibitem{lohmiller1998}
W.~Lohmiller and J.-J.~E. Slotine, ``On {{Contraction Analysis}} for
  {{Non-linear Systems}},'' {\em Automatica}, vol.~34, pp.~683--696, June 1998.

\bibitem{lee2020a}
J.~G. Lee and H.~Shim, ``A tool for analysis and synthesis of heterogeneous
  multi-agent systems under rank-deficient coupling,'' {\em Automatica},
  vol.~117, p.~108952, 2020.

\bibitem{lee2018}
J.~G. Lee and H.~Shim, ``Heterogeneous Van der Pol oscillators under strong coupling,''
in {\em Proc.~of IEEE Conf.~on Decision and Control}, Miami Beach, USA, 2018.

\bibitem{lee2022}
S.~Lee and H.~Shim, ``Blended dynamics approach to distributed optimization:
  {{Sum}} convexity and convergence rate,'' {\em Automatica}, vol.~141,
  p.~110290, 2022.

\bibitem{kim2020}
T.~Kim, C.~Lee, and H.~Shim, ``Completely {{Decentralized Design}} of
  {{Distributed Observer}} for {{Linear Systems}},'' {\em IEEE Transactions on
  Automatic Control}, vol.~65, no.~11, 2020.

\bibitem{lee2022b}
J.~G. Lee and H.~Shim, ``Design of {{Heterogeneous Multi-agent System}} for
  {{Distributed Computation}},'' in {\em Trends in {{Nonlinear}} and {{Adaptive
  Control}}} (Z.-P. Jiang, C.~Prieur, and A.~Astolfi, eds.), vol.~488,
  pp.~83--108, Cham: Springer International Publishing, 2022.

\bibitem{Azzi2017}
A.~Azzi, J.~Evans, T.~Leise, J.~Myung, T.~Takumi, A.~Davidson, and S.~Brown, 
``Network dynamics mediate circadian clock plasticity,''
{\em Neuron}, vol.~93, pp.~441--450, 2017.

\bibitem{anderson1977}
B.~Anderson, ``Exponential stability of linear equations arising in adaptive
  identification,'' {\em IEEE Transactions on Automatic Control}, vol.~22,
  pp.~83--88, Feb. 1977.

\bibitem{moreno2021}
J.G.~Rueda-Escobedo and J.A.~Moreno, ``Strong Lyapunov functions for two classical problems in adaptive control,''
{\em Automatica}, vol.~124, p.~109250, 2021.

\bibitem{Khalil2002}
H.~Khalil, {\em Nonlinear {{Systems}}}.
\newblock Prentice Hall, third edition~ed., 2002.

\bibitem{appendix}
Extended version of this paper, \url{arXiv:2509.00801}

\bibitem{github}
\url{https://github.com/hyungbo/vfconsensus/tree/main/toy_example}

\end{thebibliography}
\end{document}